\makeatletter
\def\input@path{{latex/}}
\makeatother

\documentclass[12pt]{iopart}

\pdfoutput=1

\usepackage[utf8]{inputenc}
\usepackage[english]{babel}
\usepackage[LGR, T1]{fontenc}
\usepackage{amsmath}
\usepackage{float}
\usepackage{graphicx}
\usepackage{amsfonts}
\usepackage{amsthm}
\usepackage{amssymb}
\usepackage{etoolbox}
\usepackage[dvipsnames]{xcolor}
\usepackage{hyperref}
\usepackage{soul}
\usepackage{bbm}
\usepackage{commath}
\usepackage{enumerate}
\usepackage{mathtools}
\DeclareMathAlphabet{\pazocal}{OMS}{zplm}{m}{n}
\usepackage{upgreek}
\usepackage{calc}
\usepackage{accents}
\usepackage{siunitx}
\usepackage{color}
\usepackage{textcomp}
\usepackage{bm}
\usepackage{dsfont}
\usepackage{relsize}
\usepackage{braket}
\usepackage{enumitem}   
\usepackage{mathrsfs}
\usepackage[symbol]{footmisc}
\usepackage{comment}

\usepackage[numbers, sort&compress]{natbib}

\graphicspath{{fig/}}

\newcommand{\declarebsfgreek}[2]{
  \protected\csdef{bsf#1}{\mathord{\text{\bsfgreekfont#2}}}
}
\newcommand{\bsfgreekfont}{\usefont{LGR}{cmss}{bx}{it}}
\declarebsfgreek{alpha}{a}
\declarebsfgreek{beta}{b}
\declarebsfgreek{gamma}{g}
\declarebsfgreek{delta}{d}
\declarebsfgreek{epsilon}{e}
\declarebsfgreek{zeta}{z}
\declarebsfgreek{eta}{h}
\declarebsfgreek{theta}{j}
\declarebsfgreek{iota}{i}
\declarebsfgreek{kappa}{k}
\declarebsfgreek{lambda}{l}
\declarebsfgreek{mu}{m}
\declarebsfgreek{nu}{n}
\declarebsfgreek{xi}{x}
\declarebsfgreek{omicron}{o}
\declarebsfgreek{pi}{p}
\declarebsfgreek{rho}{r}
\declarebsfgreek{sigma}{s}
\declarebsfgreek{tau}{t}
\declarebsfgreek{upsilon}{u}
\declarebsfgreek{phi}{f}
\declarebsfgreek{chi}{q}
\declarebsfgreek{psi}{u}
\declarebsfgreek{omega}{w}

\DeclareMathAlphabet{\pazocal}{OMS}{zplm}{m}{n}

\newcommand{\comments}[1]{}
\newcommand{\bea}{\begin{eqnarray}}
\newcommand{\eea}{\end{eqnarray}}

\DeclareMathOperator{\sech}{sech}

\DeclareMathOperator{\erfc}{erfc}

\usepackage{environ}
\makeatletter
\NewEnviron{quantifiedequation}[1]{
  \begin{equation}
  \expandafter\make@quantifiedequation\expandafter{\BODY}{#1}
  \end{equation}
}
\newcommand{\make@quantifiedequation}[2]{
  \m@th
  \sbox\z@{$\qquad\qquad\displaystyle#2$}
  \sbox\tw@{\let\label\@gobble$\displaystyle#1$}
  \ifdim\dimexpr 1em+\wd\z@+0.5\wd\tw@+2em>0.5\displaywidth
    #2\qquad#1
  \else
    \makebox[0pt][r]{
      \makebox[\dimexpr0.5\displaywidth-0.5\wd\tw@][l]{\quad\box\z@}
    }#1
  \fi
}
\makeatother

\newcommand{\ba}{\begin{align}}
\newcommand{\ea}{\end{align}}

\makeatletter
\DeclareFontFamily{OMX}{MnSymbolE}{}
\DeclareSymbolFont{MnLargeSymbols}{OMX}{MnSymbolE}{m}{n}
\SetSymbolFont{MnLargeSymbols}{bold}{OMX}{MnSymbolE}{b}{n}
\DeclareFontShape{OMX}{MnSymbolE}{m}{n}{
	<-6>  MnSymbolE5
	<6-7>  MnSymbolE6
	<7-8>  MnSymbolE7
	<8-9>  MnSymbolE8
	<9-10> MnSymbolE9
	<10-12> MnSymbolE10
	<12->   MnSymbolE12
}{}
\DeclareFontShape{OMX}{MnSymbolE}{b}{n}{
	<-6>  MnSymbolE-Bold5
	<6-7>  MnSymbolE-Bold6
	<7-8>  MnSymbolE-Bold7
	<8-9>  MnSymbolE-Bold8
	<9-10> MnSymbolE-Bold9
	<10-12> MnSymbolE-Bold10
	<12->   MnSymbolE-Bold12
}{}

\newcommand{\ignore}[1]{}
\newcommand{\nobibentry}[1]{{\let\nocite\ignore\bibentry{#1}}}


\begin{document}

\title[Thermal-fluctuator-driven decoherence of an oscillator]
{Thermal-fluctuator-driven decoherence of an oscillator near-resonantly coupled to a two-level system}

\setcounter{footnote}{0}
\author{Thomas J. Antolin\,\footnote{Corresponding author}, Jonas Glatthard, and Andrew D. Armour}
 
\address{School of Physics and Astronomy and Centre for the Mathematics and Theoretical Physics of Quantum Non-Equilibrium Systems, University of Nottingham, Nottingham NG7 2RD, UK}

\ead{pcytja@nottingham.ac.uk}

\begin{abstract}
Recent experiments on a range of engineered quantum systems have highlighted the important role of interacting two-level systems (TLSs) in modifying device properties and generating fluctuations. Focusing on the case of an oscillator coupled to a single near-resonant TLS, we explore how interactions between the TLS and lower-frequency thermally activated two-level fluctuators (TLFs) degrade the oscillator's coherence. Depending on the strength of the couplings, a single TLF can give rise to coherence oscillations that appear alongside, or supplant, Rabi oscillations of the oscillator--TLS system. Bath-driven transitions in the TLF cause irreversible coherence decay at a rate that is highly sensitive to both the couplings and the transition rate. For an ensemble of TLFs, we identify and characterise the different regimes of non-exponential phase-averaging-driven coherence decay that the oscillator can display. Using numerical calculations, we examine the extent to which systems with just a few TLFs differ from the limit of a large (continuum) TLF ensemble. Our work provides a theoretical framework for understanding the interplay of coherent TLS interactions and TLF-induced dephasing in quantum devices such as superconducting and phononic resonators.

\end{abstract}

\section{Introduction}
The fascinating properties of two-level systems (TLSs) have been studied extensively across a wide range of systems, from disordered solids to engineered quantum devices. Early work inferred the properties of TLS ensembles within amorphous solids through the changes in frequency and damping of acoustic modes that they induce\,\cite{phillips1972,anderson1972,phillips1987,hunklinger1988,stephens2021}. Over the last two decades, researchers have investigated the significant dephasing that TLSs give rise to in superconducting qubits\,\cite{paladino2002,galperin2006,bergli2009,paladino2014,matityahu2016,muller2019}, whilst also finding ways to use the qubits to probe the coherence of individual TLSs\,\cite{simmonds2004,lisenfeld2015,muller2015,muller2019}. Significant attention has been devoted to understanding not just the average effect that TLSs have on superconducting devices, but also the corresponding temporal fluctuations\,\cite{burnett2014,paladino2014,faoro2015,Bejanin2021}. In the last few years, experiments have also started to uncover the impact of TLSs on the properties of mechanical resonators operating in the quantum regime\,\cite{cattiaux2021,cleland2024,bozkurt2025,yuksel2025}.

Initial theoretical efforts to describe the impact of TLSs on acoustic properties in disordered solids led to the formulation of the standard tunnelling model (STM)\,\cite{anderson1972,phillips1972,phillips1987,hunklinger1988,stephens2021}. The STM assumes a very broad distribution of non-interacting TLSs, whose effects on a given acoustic mode are calculated semiclassically. The STM has proved able to account for a wide range of observations. Key elements of the STM have been adapted and applied to understand the dephasing of superconducting qubits. The low-frequency noise that degrades the coherence of superconducting qubits often has a $1/f$ character, typically attributed to a broad distribution of TLS relaxation times\,\cite{paladino2014}. As qubits operate at a temperature well below their splittings, $E\gg k_{\rm{B}}T$, attention focused on those TLSs with much lower splittings, which remain subject to thermal fluctuations, known as two-level fluctuators (TLFs). Incorporating the assumptions of a broad distribution of TLF properties and neglecting inter-TLF interactions led to a celebrated model of qubit dephasing capable of describing non-Gaussian behaviour\,\cite{paladino2002,galperin2006,bergli2009,paladino2014,matityahu2016}. Most recently, a variant of this `spin--fluctuator' model has been applied to describe the dephasing of superpositions of the two lowest Fock states of a mechanical (phononic crystal) resonator\,\cite{bozkurt2025}. 

Despite the initial success of the STM, its assumption of non-interacting TLSs is now widely regarded as insufficient to fully account for experiments on either amorphous solids or superconducting or nanomechanical quantum devices\,\cite{burnett2014, faoro2015,muller2015,carruzzo2020,cattiaux2021,maksymowych2025,yuksel2025}. This has led to the development of models in which a structured bath of TLSs is assumed, where higher-energy TLSs undergo frequency noise due to coupling to TLFs. A notable success of this structured TLS-bath approach was the development of a semiclassical model capable of accounting for the strongly counterintuitive temperature dependence observed in the frequency fluctuations of superconducting resonators\,\cite{burnett2014,faoro2015}.

In very recent experiments\,\cite{yuksel2025}, individual TLSs have been tuned into resonance with a phononic crystal resonator in a regime where their thermal fluctuations are negligible. The telegraph noise seen in these\,\cite{yuksel2025} and similar experiments\,\cite{maksymowych2025} is thought to arise from a small number of thermally driven TLFs. Inspired by these studies, we address the question of how TLFs affect the coherence of  a superconducting or mechanical oscillator near-resonantly coupled to an individual TLS. Since the small mode volumes of phononic crystal resonators mean that the number of TLFs can be small\,\cite{bozkurt2025,maksymowych2025,yuksel2025}, we begin by considering a single TLF and then explore when and how behaviour associated with a continuum distribution emerges.

Adopting a simple theoretical model, we obtain approximate analytic expressions describing the behaviour in a wide range of different parameter regimes. Coupling of the oscillator--TLS system to a single thermalised TLF leads to additional periodic oscillations in the oscillator coherence, beyond the oscillator--TLS Rabi oscillations. The additional oscillations act as an envelope atop the Rabi oscillations when the TLS--TLF coupling is weak, but for strong coupling the Rabi oscillations are washed out. Including dissipation, in the form of bath-induced transitions between TLF states, leads to exponential decay of the coherence with a rate that is very sensitive to the relative sizes of the different parameters.

For an ensemble of many TLFs, multiple oscillating components lead to rapid phase averaging and consequently the coherence decays even in the absence of dissipation. We uncover the regimes of non-exponential coherence decay corresponding to different relative TLS--oscillator and TLS--TLF coupling strengths for the continuum limit of a dense TLF ensemble. Numerical calculations demonstrate that the results obtained for a continuum distribution can provide a surprisingly useful description of coherence decay in even small ensembles of TLFs, especially for the initial parts of the evolution. However, situations where one TLF couples much more strongly than the rest give rise to clear signatures at later times, such as strong coherence revivals and less frequency mixing.

The rest of this paper is organised as follows. In section \ref{section: oscillator--TLS system coupled to a bath of fluctuators}, we introduce our model oscillator--TLS--TLF system and establish how Rabi oscillations between oscillator and TLS give rise to an apparent loss of coherence. In section \ref{section: single fluctuator} we explore the behaviour for a single fluctuator, looking at how the coherence dynamics depends on the relationship between the oscillator--TLS and TLS--TLF coupling strengths, as well as the effect of dissipation, which manifests as transitions between TLF states due to a thermal bath. In section \ref{section: ensemble of fluctuators}, we analyse the behaviour associated with a broad distribution of TLFs and use numerical calculations to investigate how this emerges as the number of TLFs is increased. Finally, we draw our conclusions and discuss some of the ways in which our work might be developed in section \ref{section: conclusions and discussion}.

\section{Oscillator--TLS system coupled to a set of fluctuators}
\label{section: oscillator--TLS system coupled to a bath of fluctuators}

\subsection{Model Hamiltonian}

Our starting point is a model which incorporates both coherent coupling between a harmonic mode and a near-resonant TLS, and dispersive coupling between the TLS and a set of $N$ lower-frequency TLFs. The Hamiltonian takes the form
\begin{equation}
    \hat{H} = \omega_0\hat{a}^\dagger\hat{a} + \frac{\varepsilon_T}{2}\hat{\sigma}_z + g(\hat{a}\hat{\sigma}_+ + \hat{a}^\dagger\hat{\sigma}_-) + \sum_{j=1}^N\left(\frac{\varepsilon_j}{2}\hat{\tau}_{z,j} + \lambda_j\hat{\sigma}_z\hat{\tau}_{z,j}\right),
\label{equation: hamiltonian}
\end{equation}
where $\omega_0$ is the bare frequency of the oscillator with lowering (raising) operator $\hat{a}$ ($\hat{a}^\dagger$) and $\varepsilon_T$ the TLS energy splitting with longitudinal spin operator $\hat{\sigma}_z$ and spin lowering (raising) operator $\hat{\sigma}_-$ ($\hat{\sigma}_+$). The $j$-th TLF has energy splitting $\varepsilon_j$ and longitudinal spin operator $\hat{\tau}_{z,j}$. The TLS--oscillator coupling is $g$ and $\lambda_j$ is the coupling of the $j$-th TLF to the TLS. Generalisations to include couplings to multiple TLSs with a range of frequencies and to include direct oscillator--TLF couplings are relatively straightforward, but we do not pursue them here and instead focus on the impact of TLFs on the coherence of the oscillator--TLS system. A related model focusing on qubit dephasing was explored in Ref.\,\cite{Matityahu2024}.

Underlying our model\,\footnote{See \ref{app: TLS properties} for a brief summary of the microscopic properties of TLSs in disordered materials.} and underpinning  our inclusion of a purely dispersive TLS--TLF interaction\,\cite{muller2015, faoro2015, matityahu2016, maksymowych2025}, is an assumed separation of energy scales\,\cite{bergli2009,burnett2014,faoro2015,matityahu2016,maksymowych2025}. This distinguishes the oscillator and TLS, which are unaffected by thermal noise, from the lower energy TLFs: $\varepsilon_T\sim\omega_0\gg k_{\rm{B}}T\gg\varepsilon_j$, where $T$ is the temperature of the surroundings.  
Consistent with this, we use the well-known Jaynes--Cummings model\,\cite{jaynes1963, garrison2008, yuksel2025} to describe the oscillator--TLS interaction, assuming the oscillator and TLS are also weakly coupled, $g\ll\min(\varepsilon_T,\omega_0)$\,\cite{wang2015}. Finally, our focus here is on how coupling to TLFs affects the near-resonant TLS, hence we have neglected interactions between the TLFs\,\cite{bergli2009,faoro2015,matityahu2016}.

\subsection{Initial states}

The decoherence of a given system can be quantified by preparing it in an initial (pure) superposition state and calculating how the corresponding coherence decays over time. We assume that only the TLFs are affected by thermal noise; hence we consider an initial state with the oscillator in a pure state $\ket{\psi(0)}$, the TLS in its ground state $\ket{g}$ and each TLF in a thermal state
\begin{equation}
    \hat{\varrho}(0) = \ket{\Psi(0)}\bra{\Psi(0)}\otimes\bigotimes^N_{j=1}\left(p_{+,j}\ket{+}_j\bra{+}_j + p_{-,j}\ket{-}_j\bra{-}_j\right),
\label{equation: initial state}
\end{equation}
where $\ket{\Psi(0)} = \ket{\psi(0)}\otimes\ket{g}$ is the initial state of the oscillator--TLS (JC) subsystem, and
\begin{equation}
    p_{\pm,j} = \frac{1}{2}\left[1\mp\tanh\left(\frac{\varepsilon_j}{2k_{\rm{B}}T}\right)\right],
\end{equation}
is the population of the $\ket{\pm}_j$ state corresponding to the $j$-th TLF. An obvious limit to consider is the case where\,\cite{bergli2009}  $k_{\rm{B}}T/\varepsilon_j\rightarrow+\infty$, which we shall refer to as the scale-separated regime. In this limit, the fluctuators are effectively thermally saturated, such that $p_{\pm,j}=1/2$ for all $j$, leading to a considerable simplification of the analysis.

Unlike the case of a qubit, there are numerous interesting initial oscillator states to explore\,\cite{remus2009,remus2012}. For example, a simple Fock-state superposition\,\cite{bozkurt2025}, $\ket{\psi(0)} \propto \ket{0} + \ket{1}$, a coherent state\,\cite{cleland2024} $\ket{\psi(0)} =\ket{\alpha}$, with amplitude $\alpha$, and  superpositions of coherent states (Schr\"{o}dinger cat states)\,\cite{walls1985, remus2009}. Here, we will restrict ourselves to the simplest relevant initial state $\ket{\psi_{01}(0)} = c_0\ket{0} + c_1\ket{1}$, with probability amplitudes $c_0,c_1$. Although this state does not allow us to explore all of the rich possibilities for an oscillator, its behaviour is already interesting enough to warrant careful investigation. More complex states are left for future studies. 

\subsection{Coherence dynamics of the oscillator--TLS system}

Before looking at the effect of the TLFs, we first review the dynamics of the closed oscillator--TLS (Jaynes--Cummings) system. Although this involves fully coherent Rabi oscillations between the oscillator and the TLS, at short times it leads to an apparent loss of oscillator coherence.

Throughout this work, we quantify the coherence using
\begin{equation}
    C(t) \equiv \left|\frac{\braket{\hat{a}(t)}}{\braket{\hat{a}(0)}}\right|.
\label{equation: coherence measure}
\end{equation}
This measure is precisely the normalised magnitude of the oscillator coherence for an initial state $\ket{\psi(0)}\propto\ket{0}+\ket{1}$, since in this subspace $\braket{\hat a(t)}=\varrho_{10}(t)$. $C(t)$ is also useful for initial coherent states as it is sensitive to fluctuations in phase as well as dissipation.

We can express the initial state of the oscillator--TLS system, $\ket{\Psi(0)} = \ket{\psi(0)}\otimes\ket{g}$, as a sum over the corresponding Jaynes--Cummings (JC) eigenstates
\begin{equation}
    \ket{\Psi(0)} =  c_0\ket{\mathcal{G}} + c_1\left(\cos\vartheta_+\ket{1+} + \cos\vartheta_-\ket{1-}\right),
\label{equation: jc initial state}
\end{equation}
where $\ket{\mathcal{G}} = \ket{0}\otimes\ket{g}$ is the ground state and
\begin{equation}
    \ket{1\pm} = \cos\vartheta_{\pm}\ket{1}\otimes\ket{g} \pm \sin\vartheta_{\pm}\ket{0}\otimes\ket{e},
\end{equation}
are the first two excited states,  with coefficients
\begin{equation}
    \cos\vartheta_{\pm} = \sqrt{\frac{\Omega \mp \delta}{2\Omega}}, \hspace{0.4cm} \sin\vartheta_{\pm} = \sqrt{\frac{\Omega \pm \delta}{2\Omega}},
\end{equation}
and energy splitting $\Omega = \sqrt{4g^2 + \delta^2}$, where $\delta = \varepsilon_T - \omega_0$ is the TLS--oscillator detuning. The corresponding eigenfrequencies are 
\begin{equation}
    \omega_{1\pm} = \frac{\omega_0 \pm \Omega}{2},
\end{equation}
for the first two excited states and $\omega_\mathcal{G} = -\varepsilon_T/2$ for the ground state.

For the initial state $\ket{\Psi(0)} = \ket{\psi_{01}(0)}\otimes\ket{g}$, equation (\ref{equation: jc initial state}), we obtain
\begin{equation}
    \braket{\hat{a}(t)} = c_0^*c_1\sum_{\alpha=\pm} \cos^2\vartheta_\alpha\,\mathrm{e}^{-i\omega_{\alpha}t},
\label{equation: generic initial state jc coherence}
\end{equation}
at time $t$, where $\omega_\alpha \equiv \omega_{1\alpha} - \omega_\mathcal{G} = (\varepsilon_T +\omega_0)/2 + \alpha\Omega/2$.

This leads to the coherence measure, equation (\ref{equation: coherence measure}), for the closed JC system
\begin{equation}
    C_\text{GR}(\Omega,t) = \sqrt{\cos^2\left(\frac{\Omega t}{2}\right) + \frac{\delta^2}{\Omega^2}\sin^2\left(\frac{\Omega t}{2}\right)},
\label{equation: jaynes--cummings dynamics}
\end{equation}
which we denote $\text{GR}$ to indicate the generalised Rabi oscillation of the oscillator and TLS.  At resonance, $\delta = 0$, the coherence undergoes full Rabi oscillations at frequency $g$: $C_\text{GR}(\Omega,t) = |\cos gt|$. At short times $\Omega t/2\ll 1$, we have the universal behaviour
\begin{equation}
    C_\text{GR}(\Omega,t) \simeq 1 - \frac{g^2t^2}{2},
\end{equation}
a quadratic short-time decay with characteristic frequency $g$, independent of the  detuning $\delta$.

In the following sections, we investigate the coherence of the oscillator, starting with the simplest possible case and adding elements step by step. Hence we start with  a single TLF (section \ref{section: single fluctuator}), before going on to consider  an ensemble of TLFs (section \ref{section: ensemble of fluctuators}).

\section{Single fluctuator}
\label{section: single fluctuator}

We now consider a single TLF with splitting $\varepsilon$ and coupling $\lambda$. The TLF generates additional frequencies in the dynamics. For weak TLS--TLF coupling these add a periodic envelope to the Rabi oscillations, but the Rabi oscillations instead become washed out when the coupling is strong. Including a non-zero rate for bath-induced transitions between TLF states leads to irreversible exponential decay. We begin by exploring the non-dissipative dynamics before going on to examine how thermal bath-induced TLF transitions affect the coherence.

\subsection{Non-dissipative fluctuator}

For a single TLF, we need to generalise our expression for the coherence, equation (\ref{equation: jaynes--cummings dynamics}), to sum over four rather than two frequency components, taking into account the different JC detunings arising from the two TLF eigenstates with appropriate thermal weights. We thus obtain
\begin{equation}
    C(t) = \Bigg|\sum_{\alpha=\pm}\,\sum_{\beta=\pm}p_\alpha\cos^2\left[\vartheta_\beta(\alpha)\right]\exp\left[-it\left(\alpha\lambda + \frac{\beta\Omega(\alpha)}{2}\right)\right]\Bigg|,
\label{equation: single tlf jc coherence}
\end{equation}
where $\Omega(\alpha)$ and $\vartheta_\beta(\alpha)$ are the JC Rabi frequency and mixing angle taking into account the TLF-induced frequency shift with sign $\alpha$, obtained by replacing $\delta\to\delta + 2\alpha\lambda$ in each. We proceed by exploring two important limiting cases. Firstly, weak coupling to the TLF, where the oscillator--TLS coupling dominates over the TLS--TLF coupling $g\gg|\lambda|$, and then the opposite strongly coupled fluctuator regime $g\ll|\lambda|$.

\subsubsection{Weakly coupled fluctuator:}
\label{subsection: weakly coupled non-dissipative fluctuator}

\begin{figure}[t!]
\centering
\includegraphics[width = 0.96\linewidth]{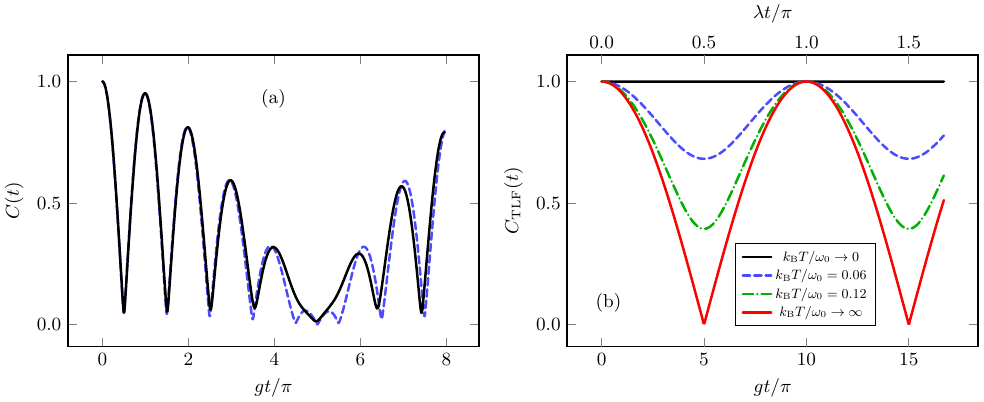}
\caption{(a) Evolution of the oscillator coherence $C(t)$ for a single weakly coupled TLF in the scale-separated limit ($p_\pm = 1/2$). Different curves compare the full expression (equation (\ref{equation: single tlf jc coherence}), black solid line) and the weak-coupling approximation (equation (\ref{equation: weakly coupled single tlf approximation}), blue dashed line). (b) Evolution of the weak-coupling TLF-induced envelope $C_\text{TLF}(t)$, obtained using equation (\ref{equation: weakly coupled single tlf approximation}), for different temperatures. We set $\varepsilon_T/\omega_0 = 1.01$, $\varepsilon/\omega_0 = 0.1$, $g/\omega_0 = 0.1$, and $\lambda/\omega_0 = 0.01$.}
\label{figure 1}
\end{figure}

When $g\gg|\lambda|$, the TLF does not affect the internal JC system strongly. In this regime, provided the system is sufficiently close to resonance, $|\delta|\ll4g^2/|\lambda|$, we can approximate the modified Rabi frequency and the modified mixing angle as their unmodified counterparts: $\Omega(\alpha) \simeq \Omega$ and $\vartheta_{\beta}(\alpha) \simeq \vartheta_{\beta}$. Under this approximation, the coherent Rabi oscillations from the two TLF configurations are identical and can be factored
\begin{equation}
    C(t) \simeq \Bigg|\sum_{\beta=\pm}\cos^2\vartheta_{\beta}\,\mathrm{e}^{-i\beta\Omega t/2}\sum_{\alpha=\pm}p_\alpha\,\mathrm{e}^{-i\alpha\lambda t}\Bigg|.
\end{equation}
Hence the TLF modulates the Rabi oscillations in the coherence via a slowly evolving envelope
\begin{equation}
    C(t) \simeq C_\text{GR}(\Omega,t)C_\text{TLF}(t)=C_\text{GR}(\Omega,t)\sqrt{\cos^2\lambda t + \tanh^2\left(\frac{\varepsilon}{2k_{\rm{B}}T}\right)\sin^2\lambda t}.
\label{equation: weakly coupled single tlf approximation}
\end{equation}
In the scale-separated limit where the temperature is much higher than the TLF splitting, $k_{\rm{B}}T/\varepsilon\rightarrow\infty$, $C(t) \simeq C_\text{GR}(\Omega,t)|\cos\lambda t|$.
 
Figure \ref{figure 1} shows that our approximate solution, equation (\ref{equation: weakly coupled single tlf approximation}), captures the exact dynamics, equation (\ref{equation: single tlf jc coherence}), well for sufficiently large $g/|\lambda|$, though a small residual frequency difference becomes apparent at longer times. As the temperature is reduced, differences in weightings of the two TLF states grow and the envelope, $C_\text{TLF}(t)$, gets increasingly shallow [figure \ref{figure 1}(b)].

The TLF-induced envelope, $C_\text{TLF}(t)$, is equivalent to the coherence modulation in the spin--fluctuator model\,\cite{galperin2006,bergli2009,matityahu2016}. This is because in both cases---the spin--fluctuator model and the weak TLF-coupling limit of our JC--fluctuator model---the internal dynamics (of the spin and JC system, respectively) are unaffected by the TLF.

\subsubsection{Strongly coupled fluctuator:}

\begin{figure}[t!]
\centering
\includegraphics[width = 0.96\linewidth]{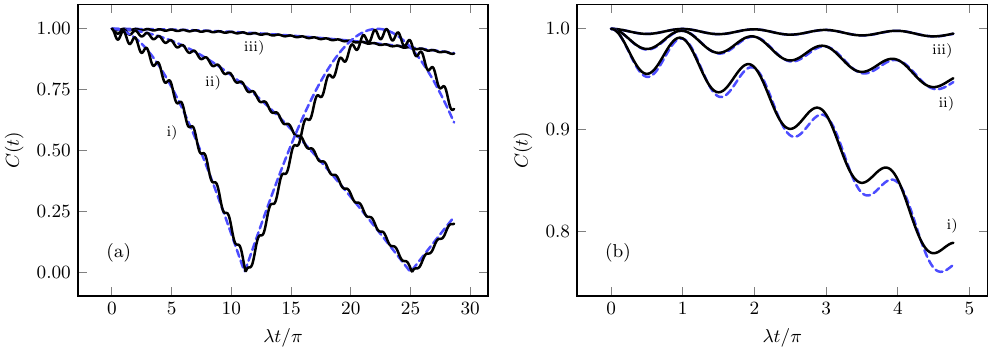}
\caption{Evolution of $C(t)$ with a single strongly coupled non-dissipative fluctuator in the scale-separated limit ($p_{\pm}=1/2$) with: i) $g/\lambda = 0.3$, ii) $g/\lambda = 0.2$ and iii) $g/\lambda = 0.1$. (a) Comparison of the full expression (equation (\ref{equation: single tlf jc coherence}), black solid lines), and the approximate expression for the strongly-coupled TLF envelope  (equation (\ref{equation: strongly coupled single tlf approximation at leading order}), blue dashed lines). (b) Behaviour over a narrower time window. Here, the higher-order expression (equation (\ref{equation: strongly coupled single tlf approximation at higher order}), blue dashed lines) is shown. In each case, we set $\delta = 0$ and $\lambda/\omega_0 = 0.1$.}
\label{figure 2}
\end{figure}

We now consider the opposite case, where the TLS--TLF coupling dominates over the oscillator--TLS coupling\,\footnote{This regime has been accessed in recent experiments\,\cite{maksymowych2025}. See also \ref{app: TLS properties}.}: $|\lambda|\gg g$. For simplicity, we focus on the on-resonance regime, $\delta = 0$, and approximate the modified generalised Rabi frequency as $\Omega\simeq 2|\lambda| + g^2/|\lambda|$. Hence in this case we find
\begin{equation}
    C(t) \simeq \sqrt{\cos^2\left(\frac{g^2t}{2\lambda}\right) + \tanh^2\left(\frac{\varepsilon}{2k_{\rm{B}}T}\right)\sin^2\left(\frac{g^2t}{2\lambda}\right)}.
\label{equation: strongly coupled single tlf approximation at leading order}
\end{equation}
As figure \ref{figure 2} shows, this describes a slow periodic envelope oscillating between unity and $\tanh(\varepsilon/2k_{\rm{B}}T)$ at frequency $g^2/2|\lambda|$. Once again, the role of temperature is to squeeze the amplitude of the envelope: in the scale-separated limit ($k_{\rm{B}}T/\varepsilon\to\infty$) the coherence envelope oscillates between unity and zero, whereas at zero temperature the envelope remains fixed at unity.

The full solution, equation (\ref{equation: single tlf jc coherence}), also displays shallow Rabi oscillations atop the much larger periodic envelope. These are not captured in equation (\ref{equation: strongly coupled single tlf approximation at leading order}). As discussed in \ref{appendix: higher-order approximate solution for a single non-dissipating, strongly-coupled fluctuator}, a higher-order approximation that includes terms $\mathcal{O}(g^2/\lambda^2)$ provides a good description of the fast Rabi oscillations with frequency $2|\lambda|$ [see figure \ref{figure 2}(b)].

Figure \ref{figure 2} also shows that as $g/|\lambda|$ is reduced and we move further into the strongly-coupled TLF regime, the TLF-induced coherence modulation actually gets weaker. This can be understood in terms of the ever larger TLS--oscillator detuning that is generated as $|\lambda|$ increases. This increases the effective energy barrier between the oscillator and TLS. In turn, this tends to insulate the oscillator from the TLS, and hence also from the influence of the TLF.

\subsection{Dissipative fluctuator}
\label{subsection: dissipative fluctuator}

We now consider the effect on the oscillator coherence of transitions between the TLF states due to the interaction with the thermal environment. For simplicity, we limit ourselves to working in the scale-separated regime, $k_{\rm{B}}T/\varepsilon\to\infty$, so that the upward and downward TLF transition rates are equal, $\gamma_+ = \gamma_- = \gamma$. Starting from a simple Lindblad master equation for the system, we derive coupled equations of motion for the expectation values needed to obtain $C(t)$. We then outline the simple approximate equations that describe the coherence dynamics in a range of different parameter regimes. Once again, we consider the cases of weakly coupled ($|\lambda|\ll g$) and strongly coupled ($|\lambda|\gg g$) TLF in turn. Even with the simplifying approximations we have made, several interesting behaviours emerge.

The master equation is derived using a `local' approach that assumes a weakly coupled TLF, $|\lambda|\ll g$, but it can also be applied in the strongly coupled regime, $g\ll|\lambda|$, provided there remains a separation of energy scales $|\lambda|\ll k_{\rm B}T\ll\varepsilon_T$ and the bath density of states varies sufficiently slowly over the relevant transition frequencies. Under these conditions, the bath-induced transition rates between the TLS--TLF eigenstates can be approximated by the common rate $\gamma$ used here. Outside this regime, the local treatment is not necessarily reliable. Further details are given in \ref{appendix:dissipation}.

\subsubsection{Weakly coupled fluctuator, $|\lambda|\ll g$:} 
In this regime the TLF is weakly coupled to the oscillator--TLS system and the latter undergoes Rabi oscillations, but within an envelope generated by the TLF which is now damped.  In this case, the master equation leads to the approximate expression
\begin{equation}
    C(t) \simeq {\mathrm{e}}^{-\gamma t}\left|\cos gt\left[\cos\eta t+\frac{\gamma}{\eta}\sin\eta t\right]\right|, \label{eq:wcfdecay}
\end{equation}
where $\eta= \sqrt{\lambda^2-\gamma^2}$. Once again, the weakly coupled TLF case is closely related to the coherence modulation of the spin--fluctuator model\,\cite{bergli2009}.

\begin{figure}[t!]
\centering
\includegraphics[width = 0.96\linewidth]{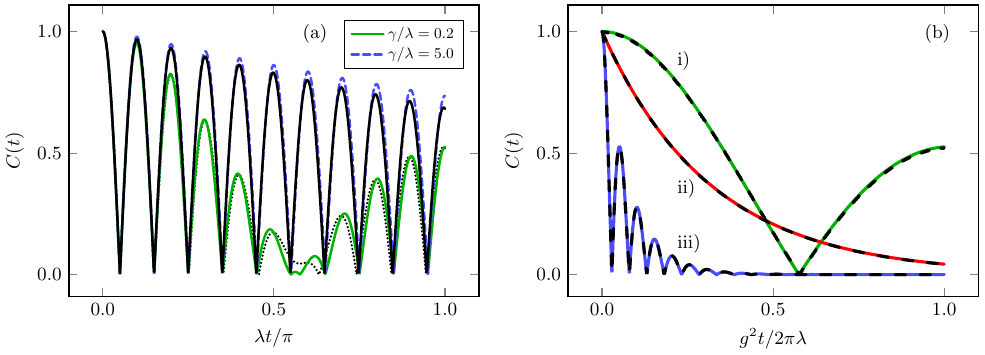}
\caption{Evolution of $C(t)$ with a single dissipative TLF in the scale-separated limit ($p_{\pm}=1/2$) for varying $\gamma/\lambda$. (a) Comparison for a weakly-coupled TLF ($g\gg|\lambda|$) of the numerically solved master equation (equation (\ref{equation: master equation}), black solid and dotted lines for
$\gamma/\lambda=0.2$ and $5.0$, respectively) and the approximate expression (equation (\ref{eq:wcfdecay}), green solid and blue dashed lines). (b) Comparison for a strongly-coupled TLF ($g\ll|\lambda|$) of the numerically solved master equation (equation (\ref{equation: master equation}), black dashed lines) and the approximate expressions for i) underdamping ($\gamma/\lambda=0.001$, green), ii) intermediate damping ($\gamma/\lambda=1$, red), and iii) overdamping ($\gamma/\lambda=50$, blue). The underdamping and overdamping approximations are given by equation (\ref{eq:weakdamp}), while the intermediate damping approximation is $C(t)\simeq \exp(-g^2\gamma t/2\lambda^2)$. In (a) we set $g/\omega_0 = 0.1$ and $\lambda/\omega_0 = 0.01$, while in (b) we set $g/\omega_0 = 0.01$ and $\lambda/\omega_0 = 0.1$, and in both we set $\delta = 0$.}
\label{figure 3}
\end{figure}

Oscillations in the Rabi envelope are underdamped when $|\lambda|\gg\gamma$, decaying at the TLF transition rate, $\gamma$. In the opposite limit, $\gamma\gg|\lambda|$, the TLF is strongly overdamped, leading to decay of the Rabi oscillation envelope at a rate $ \lambda^2/2\gamma$. For $|\lambda| \rightarrow \gamma$, so that $\eta\rightarrow 0$, there is critical damping and $C(t)\simeq\mathrm{e}^{-\gamma t}\,|\cos gt\,|(1 + \gamma t)$. In this special case, the behaviour is  non-exponential, with $C_\text{TLF}(t)\simeq\mathrm{e}^{-\gamma t}\,(1+\gamma t)$.

\subsubsection{Strongly coupled fluctuator, $|\lambda|\gg g$:}

For a strongly coupled fluctuator, three different regimes emerge depending on the relative strength of $|\lambda|/\gamma$. For weak dissipation ($|\lambda|\gg \gamma$), the coherent oscillations seen in the non-dissipative case, equation (\ref{equation: strongly coupled single tlf approximation at leading order}), become damped with
\begin{equation}
    C(t)={\rm{e}}^{-\Gamma t}\left|\cos\kappa t+\frac{\Gamma}{\kappa}\sin\kappa t\right|,
\label{eq:weakdamp}
\end{equation}
where $\kappa=\sqrt{(g^2/2\lambda)^2-\Gamma^2}$ and $\Gamma=\gamma$.
Here again, underdamping, overdamping and critical damping can be found, depending on the relative sizes of $\gamma$ and $g^2/2|\lambda|$.

For intermediate damping, where $|\lambda|\sim\gamma$, we find that the coherence envelope decays exponentially at a rate $\simeq g^2\gamma/2\lambda^2$, and there are no oscillations. Finally, for strong damping, $\gamma\gg |\lambda|$, equation (\ref{eq:weakdamp}) again holds, but now with $\kappa=\sqrt{g^2-\Gamma^2}$ and $\Gamma=\lambda^2/\gamma$. Strikingly, in this regime the effects of the strongly coupled TLF are muted, allowing Rabi oscillations to re-emerge with frequency $g$, albeit rapidly damped. 

The approximate expressions for $C(t)$ in the different dissipative regimes are compared with numerical solutions of the corresponding master equation [equation (\ref{equation: master equation})] in figure \ref{figure 3}. For both weak and strong TLF--TLS couplings, the numerical results confirm the accuracy of the approximate expressions.

\section{Ensemble of fluctuators}
\label{section: ensemble of fluctuators}

We now consider a set of $N$ TLFs, focusing on the non-dissipative regime (neglecting transitions in the TLF states), assuming that the timescales over which the TLFs fluctuate are long compared to the time taken for the coherence to decay. In this case, the thermal environment only enters through the initial thermal state of the TLF ensemble, acting as a form of frozen disorder\,\cite{matityahu2016}. Each of the different contributions in the thermal state has a slightly different oscillation frequency. Summing over the contributions leads to phase averaging\,\footnote{This is appropriate for modelling an experimental situation in which individual coherence decay measurements are made over a timescale that is short enough for the TLFs to be highly likely to remain in a fixed configuration during a given run, but repeated sufficiently often (and with sufficient delay between measurement runs) to ensure that the full thermal distribution of configurations is sampled.}, resulting in non-exponential decay of the coherence\,\cite{bergli2009,HarocheRaimond}. This situation is likely to be relevant for experiments conducted at sufficiently low temperatures: the relaxation rates of thermally active TLFs (driven by phonons in insulating structures) are strongly temperature dependent, getting slower as they are cooled\,\cite{matityahu2016} (see also the discussion in \ref{appendix: connection to microscopic properties of the ensemble}). 

For an ensemble of $N$ non-dissipating TLFs, the evolution is an average over $2^N$ TLF-modified JC Hamiltonians: one for each configuration. The JC Hamiltonian corresponding to TLF configuration $\bm{\alpha} = (\alpha_1,\alpha_2,\cdots,\alpha_N)\in\{\pm1\}^N$ is given by
\begin{equation}
    \hat{H}(\bm{\alpha}) = \omega_0\hat{a}^\dagger\hat{a} + \frac{\varepsilon_T(\bm{\alpha})}{2}\hat{\sigma}_z + g(\hat{a}\hat{\sigma}_+ + \hat{a}^\dagger\hat{\sigma}_-),
\end{equation}
where $\varepsilon_T(\bm{\alpha}) = \varepsilon_T + 2(\bm{\alpha}\cdot{\bm{\lambda}})$ is the modified TLS level-splitting, and $\bm{\lambda} = (\lambda_1,\lambda_2,\cdots,\lambda_N)$ is the set of TLS--TLF couplings embedded into the vector space $\mathbb{R}^N$ such that $\cdot$ denotes the inner (dot) product.

The reduced density operator of the oscillator--TLS system at time $t$, accounting for the TLF ensemble, is given by
\begin{equation}
    \hat{\varrho}_\text{JC}(t) = \sum_{\bm{\alpha}\in\{\pm1\}^N}P(\bm{\alpha})\hat{\varrho}_{\bm{\alpha}}(t),
\end{equation}
where $P(\bm{\alpha}) = \prod_{j=1}^Np_{\alpha_j,j}$ is the thermal population of TLF configuration $\bm{\alpha}$ and
\begin{equation}
    \hat{\varrho}_{\bm{\alpha}}(t) = \mathrm{e}^{-i\hat{H}(\bm{\alpha})t}\,\ket{\Psi(0)}\bra{\Psi(0)}\,\mathrm{e}^{i\hat{H}(\bm{\alpha})t},
\end{equation}
is the contribution to the unitary evolution from the TLF configuration encoded by $\bm{\alpha}$.

For our simple oscillator superposition initial state, equation (\ref{equation: jc initial state}), we find that
\begin{equation}
    C(t) = \left|\sum_{\bm{\alpha}\in\{\pm1\}^N}P({\bm{\alpha}})\,\mathrm{e}^{-i(\bm{\bm{\alpha}}\cdot\bm{\lambda})t}\left(\cos\left[\frac{\Omega({\bm{\alpha}})t}{2}\right] + \frac{i\delta({\bm{\alpha}})}{\Omega({\bm{\alpha}})}\sin\left[\frac{\Omega({\bm{\alpha}})t}{2}\right]\right)\right|,
\label{equation: many tlf jc coherence}
\end{equation}
where we have defined an ensemble-modified Rabi frequency $\Omega({\bm{\alpha}}) = \sqrt{4g^2 + \delta^2({\bm{\alpha}})}$ and detuning $\delta({\bm{\alpha}}) = \varepsilon_T(\bm{\alpha}) - \omega_0$. 
Solving equation (\ref{equation: many tlf jc coherence}) numerically provides a way of uncovering the effects of increasing $N$ (see figures \ref{figure 4}, \ref{figure 5}, \ref{figure 6}, and \ref{figure 7} below). However, for a large TLF ensemble it is possible to make progress by assuming the limit of a continuous distribution as we now show.

Defining the TLS--TLF coupling inner product associated with the configuration $\bm{\alpha}$ as $\Lambda = \bm{\alpha}\cdot\bm{\lambda}$, we can rewrite equation (\ref{equation: many tlf jc coherence}) as a weighted average over $\Lambda$-values
\begin{equation}
    C(t) = \Bigg|\,\sum_{\Lambda} w(\Lambda)f(\Lambda,t)\,\Bigg|,
\end{equation}
with weight function
\begin{equation}
    w(\Lambda) \equiv \sum_{\bm{\beta}\in\{\pm1\}^N}P(\bm{\beta})\,\delta_{\Lambda, \bm{\beta}\cdot\bm{\lambda}},
\end{equation}
 where $\delta_{\Lambda, \bm{\beta}\cdot\bm{\lambda}}$ denotes the Kronecker delta, and the contribution is
\begin{equation}
    f(\Lambda,t) \equiv \mathrm{e}^{-i\Lambda t}\left(\cos\left[\frac{\Omega(\Lambda) t}{2}\right] + \frac{i\delta(\Lambda)}{\Omega(\Lambda)}\sin\left[\frac{\Omega(\Lambda) t}{2}\right]\right),
\end{equation}
with $\Omega(\Lambda) = \sqrt{4g^2 + \delta^2(\Lambda)}$ and $\delta(\Lambda) = \delta + 2\Lambda$ the TLF-modified Rabi frequency and oscillator--TLS detuning associated with the inner product $\Lambda$.

For a sufficiently large ensemble, $N\gg1$, consisting of independent TLFs where no single TLF---or small set of TLFs---dominates the variance, we can invoke the central limit theorem (CLT)\,\cite{billingsley1995}, approximating the inner product distribution as Gaussian 
\begin{equation}
    w(\Lambda) \simeq \mathcal{N}(\mu,\sigma^2) = \frac{1}{\sqrt{2\pi}\sigma}\exp\left[-\frac{(\Lambda - \mu)^2}{2\sigma^2}\right],
\end{equation}
where $\mu = -\sum_{j=1}^N\lambda_j\tanh(\varepsilon_j/2k_{\rm{B}}T)$ and $\sigma^2 = \sum_{j=1}^N\lambda_j^2\sech^2(\varepsilon_j/2k_{\rm{B}}T)$ are the thermally-weighted mean and variance, respectively. Hence, in the large-ensemble limit we have
\begin{equation}
    C(t) \simeq \bigg|\int^\infty_{-\infty}d\Lambda\,\mathcal{N}(\mu,\sigma^2)f(\Lambda,t)\bigg|.
\label{equation: continuous tlf jc coherence}
\end{equation}
 Although this is only strictly correct in the large $N$ limit, where the contribution to the total variance from each individual TLF, $\sigma_j^2$, becomes vanishingly small\,\cite{billingsley1995} so that $R = \max_j(\sigma_j^2)/\sigma^2 \to 0$, we may anticipate that it still remains useful in cases where $R\ll 1$. Correlations between TLFs, arising for example from direct TLF--TLF interactions, would invalidate the assumption of independent contributions used here and, if sufficiently strong, could lead to significant deviations from the Gaussian distribution predicted by the CLT.

Connection with the microscopic details of the TLF ensemble in a specific system can be made, requiring only calculation of $\mu$ and $\sigma$, as discussed in \ref{appendix: connection to microscopic properties of the ensemble}. Since in general the distribution of TLF couplings will be even and we assume $k_{\rm{B}}T\gg\varepsilon_j$,  we can expect $\mu\simeq 0$ in the continuum limit,  simplifying things even further\,\cite{muller2015,carruzzo2020}. 
For continuum TLF-distributions a linear temperature dependence, $\sigma^2\propto T$, then follows provided the distribution over energies is uniform\,\cite{matityahu2016,muller2015}. 

To mirror what we did for a single TLF, we now examine the large-ensemble behaviour in two limits: one where the JC coupling dominates over the typical TLS--TLF coupling (a narrow TLF ensemble) and one in the opposite regime where the TLS--TLF couplings can dominate (a broad TLF ensemble). In both cases, approximate analytical expressions are obtained.

\subsection{Narrow ensemble}

Firstly, we consider a TLF ensemble which is both large, $N\gg1$, and narrow, $g\gg\sigma$, with $|\mu|\ll g$. For simplicity, we also assume zero bare oscillator--TLS detuning, $\delta = 0$. Exploiting the fact that the Gaussian inner product distribution is strongly peaked about $\mu$, we expand 
\begin{equation}
    \ln{f(\Lambda, t)} \simeq \ln{f(\mu,t)} + (\Lambda - \mu)\frac{\partial \ln{f(\Lambda,t)}}{\partial\Lambda}\bigg{|}_{\Lambda = \mu} + \mathcal{O}\left[(\Lambda - \mu)^2\right],
\end{equation}
an approach which is valid when the first-order term, $\mathcal{O}(\Lambda - \mu)$, is a small correction to the zeroth-order term, $\mathcal{O}(1)$. The first term generates a Rabi oscillation modified by the thermal mean of the TLF bath, whereas the second term accounts for the TLF fluctuations that produce an envelope decay. 

When $g\gg|\Lambda|$, the dependence of $\Omega(\Lambda)$ on $\Lambda$ is very weak and we can approximate 
\begin{equation}
    \frac{\partial \ln f(\Lambda,t)}{\partial\Lambda}\bigg{|}_{\Lambda = \mu} \simeq -it,
\end{equation}
provided we also have $t\ll g/\Lambda^2$. For typical $\Lambda$-values within the narrow distribution, $|\Lambda|\sim\sigma \ll g$, the approximation will therefore hold for times $t\ll g/\sigma^2$.

Applying the expansion to equation (\ref{equation: continuous tlf jc coherence}), we obtain
\begin{equation}
    C(t) \simeq C_\text{GR}\left(\Omega({\mu}),t\right)\,\exp\left[-\frac{t^2}{2}\sum_{j=1}^N \lambda_j^2\sech^2\left(\frac{\varepsilon_j}{2k_{\rm{B}}T}\right)\right].
\label{equation: narrow tlf short time approximation}
\end{equation}
As with the single weakly coupled TLF, the coherence factorises into a Rabi oscillation part and a decaying envelope. Note that a non-zero $\mu$ produces a change in the Rabi frequency $\Omega(\mu)$ which becomes significant at sufficiently long times. The Gaussian decay behaviour, $\exp(-\sigma^2t^2/2)$, arises from averaging over the $2^N$ phases generated by the individual TLF configurations and is reminiscent of the collapse of Rabi oscillations seen in the JC model when the oscillator is initially in a coherent state\,\cite{HarocheRaimond}. 
At longer times, $t\gtrsim g/\sigma^2$, it can be shown that the TLF envelope crosses over to an exponential decay, similar to Ref.\,\cite{matityahu2016}. However, for the narrow TLF-ensemble, this crossover only occurs after the Gaussian decay envelope has already strongly suppressed the Rabi oscillations, $C_\text{TLF}(t\sim g/\sigma^2) \sim \exp(-g^2/2\sigma^2)\ll1$.

\begin{figure}[t!]
\centering
\includegraphics[width = 0.7\linewidth]{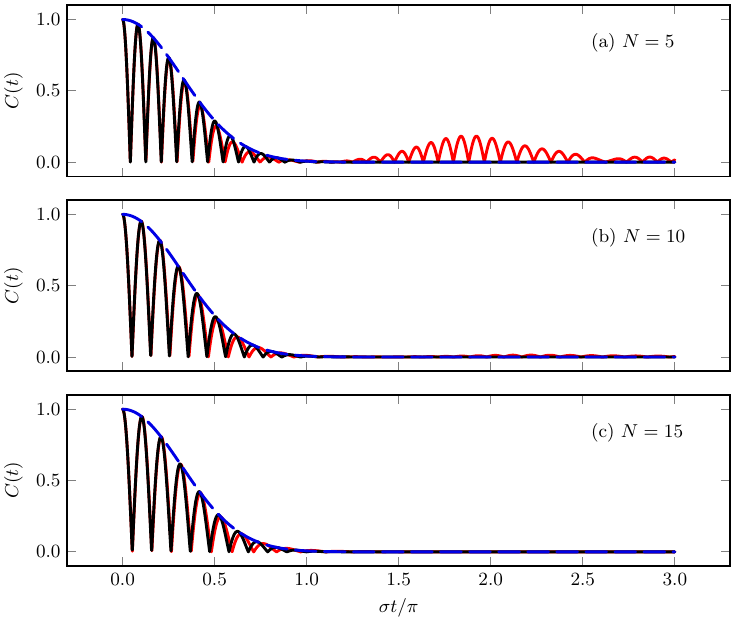}
\caption{Evolution of $C(t)$ for a narrow ensemble ($g\gg\sigma$) of (a) 5, (b) 10, and (c) 15 TLFs in the scale-separated limit ($p_{\pm,j}=1/2$, $\forall j$) with $\sigma/g=0.084, 0.102$ and $0.105$, respectively. In each case, the exact solution (equation (\ref{equation: many tlf jc coherence}), red solid lines) is compared with  the approximate expression (equation (\ref{equation: narrow tlf short time approximation}), black solid lines)  and the envelope,  $C_\text{TLF}(t) = \exp(-\sigma^2t^2/2)$ (blue dashed lines).  We set $\delta = 0$, $g/\omega_0 = 0.1$ and sampled the TLS--TLF couplings from a narrow uniform random distribution $\lambda\in[-0.05g,0.05g]$.}
\label{figure 4}
\end{figure}

\begin{figure}[t!]
\centering
\includegraphics[width = 0.7\linewidth]{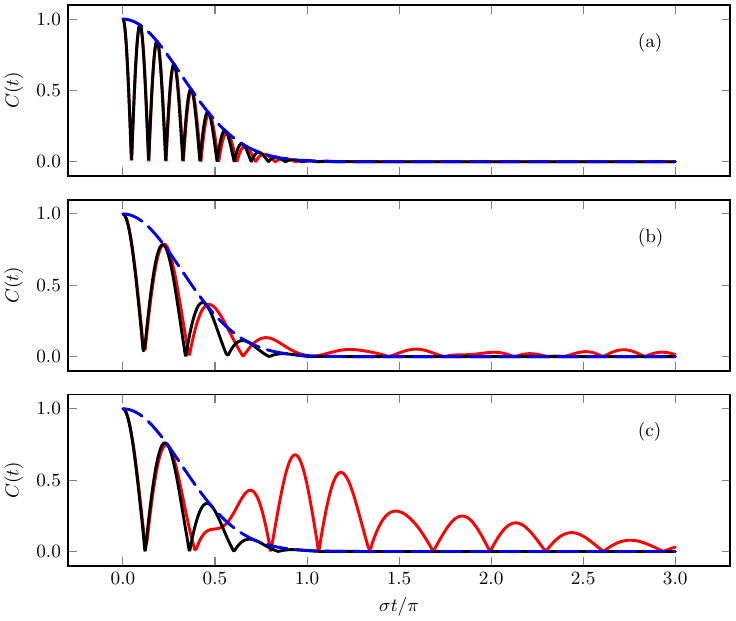}
\caption{Evolution of $C(t)$ for a narrow  ensemble ($g\gg\sigma$) of  15 TLFs in the scale-separated limit ($p_{\pm,j}=1/2$, $\forall j$). (a), (b) and (c) are three different samplings of TLS--TLF couplings from a uniform random distribution of locations in a two-dimensional substrate (see text for details), leading to $\sigma/g=0.093, 0.23$ and $0.24$, respectively. The corresponding $R = \max_j(\sigma_j^2)/\sigma^2$ values are (a) $0.22$, (b) $0.45$ and (c) $0.92$. In each case, the exact solution (equation (\ref{equation: many tlf jc coherence}), red solid lines) is compared with  the approximate expression (equation (\ref{equation: narrow tlf short time approximation}), black solid lines)  and the envelope,  $C_\text{TLF}(t) = \exp(-\sigma^2t^2/2)$ (blue dashed lines).  We set $\delta = 0$ and $g/\omega_0 = 0.1$.}
\label{figure 5}
\end{figure}

To begin testing our analytic approximation and the convergence to the large-$N$ limit, we looked at $C(t)$ for a range of different realisations with couplings drawn from a simple uniform distribution and different values of $N$. Figure \ref{figure 4} shows how well equation (\ref{equation: narrow tlf short time approximation}) can capture the exact evolution of a single realisation, even down to $N=5$ at short times. For $N=5$, there is significant sample-to-sample variation, particularly at later times. The key signature of a `large' ensemble is a decay of Rabi oscillations not immediately followed by any revival. A more extensive statistical analysis of these effects is provided in~\ref{appendix: statistical analysis of the onset of continuum-like behaviour in narrow ensembles}, where we show that continuum-like behaviour can emerge for ensembles containing as few as $6$--$8$ TLFs, although its onset depends on both the ensemble size and the particular realisation of the individual TLF coupling strengths.

One of the most significant microscopic details is the dependence of TLS--TLF couplings on the corresponding physical separation\,\cite{matityahu2016,carruzzo2020,behunin2016}, $r$, such that $\lambda\propto (r_0/r)^d$ where $d$ is the dimension of the host medium and $r_0$ a short distance cut-off (see \ref{appendix: connection to microscopic properties of the ensemble} for details). Whilst the TLFs are expected to be uniformly distributed in space, the corresponding coupling distribution will not be uniform. To illustrate this, we simulated a two-dimensional system by selecting uniform random coordinates $(x,y)\in[1,10]^2$ (in units of $r_0$) and then calculated a corresponding coupling using $\lambda=\pm g/(x^2+y^2)$, with signs selected at random with equal probability. Three samples for the case where $N=15$ are shown in figure \ref{figure 5}. As one would expect, there is significant sample-to-sample variation, and strong deviations from the large-ensemble approximations are found in cases where $R = \max_j(\sigma_j^2)/\sigma^2$ is not much less than unity. Interestingly, the early time behaviour is captured by our approximations in each case with deviations emerging later on in the form of coherence revival and oscillations with a narrow range of frequencies. A more detailed statistical analysis of the sample-to-sample variation and its dependence on $N$ and $R$ for small ensembles with random locations is included in~\ref{appendix: statistical analysis of the onset of continuum-like behaviour in narrow ensembles}.

\subsection{Broad ensemble}

Now we consider the opposite regime: a TLF ensemble which is both large, $N\gg1$, and broad, $\sigma\gg g$. Progress can be made if we assume that the typical inner product dominates over the JC coupling, $|\Lambda|\sim \sigma \gg g$. For the on-resonance case, $\delta = 0$, equation (\ref{equation: continuous tlf jc coherence}) reduces to
\begin{equation}
    C(t) \simeq \bigg|\int^\infty_{-\infty}d\Lambda\,\mathcal{N}(\mu,\sigma^2)\,\mathrm{e}^{ig^2t/2\Lambda}\bigg|.
\label{equation: broad tlf approximation}
\end{equation}
Although this integral does not admit a closed form solution, it can be used to derive an approximate analytical expression valid for short times.

\begin{figure}[t]
\centering
\includegraphics[width = 0.96\linewidth]{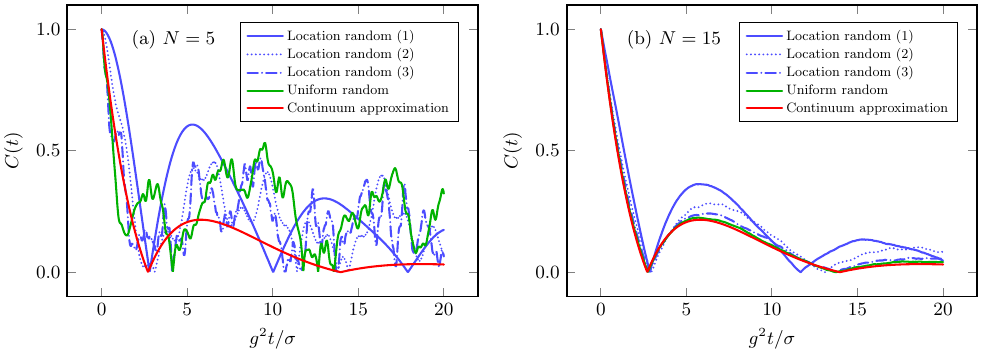}
\caption{Evolution of $C(t)$ for a broad ensemble ($\sigma\gg g$) of (a) $5$ and (b) $15$ TLFs in the scale-separated limit ($p_{\pm,j}=1/2$, $\forall j$). In each case, a single realisation of a uniform random coupling distribution (green solid line) is compared with three samples from a random distribution of locations (blue lines), details of which are given in the main text. Each is compared with the continuum approximation (equation (\ref{equation: broad tlf approximation}), red solid line). We set $\delta = 0$, $g/\omega_0 = 0.01$ and the uniform random distribution of couplings is  $\lambda\in[-5g,5g]$. In (a) $R = \max_j(\sigma_j^2)/\sigma^2$ is $0.93, 0.79, 0.53$ and $0.52$ for samples (1--3) of the random location distribution and the uniform random coupling sample, respectively. For (b) the corresponding values are $0.75, 0.60, 0.40$ and $0.18$.}
\label{figure 6}
\end{figure}

The behaviour of $C(t)$ for broad ensembles is illustrated in figure \ref{figure 6}. We compare samples from broad distributions assuming uniform random coupling and uniform random positions\,\footnote{The uniform position distribution was obtained by sampling uniform random coordinates (as for narrow ensembles discussed above) to obtain $\lambda=\pm gw/(x^2+y^2)$, though now with the scaling factor $w=200(50)$ for $N=5(15)$ to ensure the appropriate breadth is achieved ($\sigma\gg g)$.} with  our large-$N$ approximation [equation (\ref{equation: broad tlf approximation})] for $N=5$ and $N=15$ in  figures \ref{figure 6}(a) and \ref{figure 6}(b), respectively. The degree to which the continuum ensemble result is expected to apply can again be quantified via $R = \max_j(\sigma_j^2)/\sigma^2$. For $N=5$, $R\ll 1$ is never achieved, even for the sample from the uniform random couplings. Hence it is no surprise that strong sample-to-sample variations are seen. The sample from the uniform random location distribution with the largest $R$ value ($0.93$) shows that where a single coupling starts to dominate, the behaviour can look qualitatively different, even at short times, with  stronger revivals and less frequency dispersion becoming apparent at later times. For $N=15$, the samples can be well-described by the large-ensemble approximation, though there is still a significant amount of sample-to-sample variation.

\begin{figure}[t]
\centering
\includegraphics[width = 0.96\linewidth]{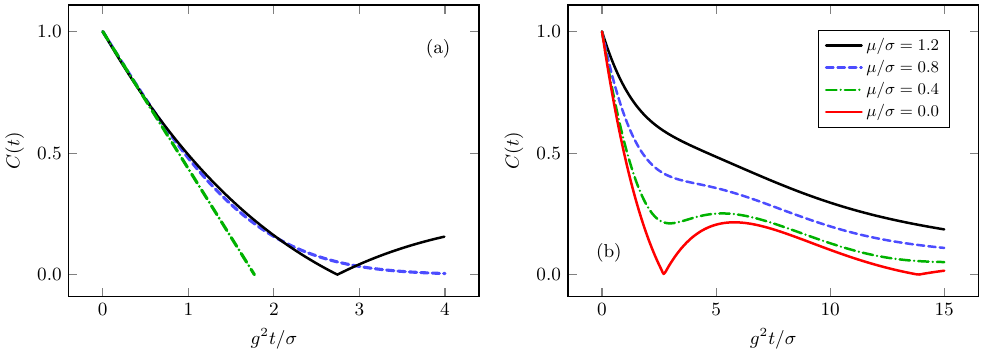}
\caption{(a) Short-time approximations to $C(t)$ in the broad-ensemble ($\sigma\gg g$)  and scale-separated ($p_{\pm,j} = 1/2$, $\forall j$) regimes. Approximations for intermediate ($t\lesssim\sigma/g^2$, equation (\ref{equation: broad distribution short time approximation}), blue dashed line) and short ($t\ll\sigma/g^2$, equation (\ref{equation: broad tlf shortest time approximation}), green dash-dotted line) times are shown together with the large-ensemble approximation (equation (\ref{equation: continuous tlf jc coherence}), black solid line). (b) $C(t)$ in the large-ensemble limit, equation (\ref{equation: continuous tlf jc coherence}), for varying $\mu/\sigma$. We set $\delta = 0$ and $g/\omega_0 = 0.01$ throughout.}
\label{figure 7}
\end{figure}

We now turn to the short time regime and focus on ensembles with $|\mu|\lesssim\sigma$, where further analytic approximations can be made. In the region where $|\Lambda|\lesssim g^2t/\sqrt{2}$, the integrand of equation (\ref{equation: broad tlf approximation}) is highly oscillatory, with a frequency that diverges as $\Lambda \rightarrow 0$. Hence, this range can be omitted\,\footnote{When $|\mu|\gtrsim\sigma$, the fast oscillating region is outside the dominant support of the distribution and our approximation strategy fails.}. Elsewhere, and for times up to $t\lesssim \sigma/g^2$, the relevant phases $g^2t/2\Lambda$ remain less than or of order unity and the integrand does not oscillate strongly, so that $\exp(ig^2t/2\Lambda) \simeq 1$. Under these conditions, we find
\begin{equation}
    C(t) \simeq \frac{1}{2}\left[\erfc\left(\frac{g^2t}{2\sigma} + \frac{\mu}{\sqrt{2}\sigma}\right) + \erfc\left(\frac{g^2t}{2\sigma} - \frac{\mu}{\sqrt{2}\sigma}\right)\right],
\label{equation: broad distribution short time approximation}
\end{equation}
where $\erfc\left(\cdot\right)$ is the complementary error function. For $|\mu|\ll\sigma$, this expression simplifies to
$ C(t) \simeq \erfc(g^2t/2\sigma)$. The phase-averaging behaviour in this case thus corresponds to an initial non-exponential decay over the characteristic timescale $\sigma/g^2$. Further simplification is possible for even shorter times, $t\ll \sigma/g^2$, with
\begin{equation}
    C(t) \simeq 1 - \frac{g^2\exp\left({-\mu^2/2\sigma^2}\right)t}{\sqrt{\pi}\sigma}.
\label{equation: broad tlf shortest time approximation}
\end{equation}

Figure \ref{figure 7}(a) shows that these short-time expressions, equations (\ref{equation: broad distribution short time approximation}) and (\ref{equation: broad tlf shortest time approximation}), match the large-ensemble expression, equation (\ref{equation: continuous tlf jc coherence}), very well in their appropriate limits. Figure \ref{figure 7}(b)  shows that when $|\mu|/\sigma$ is no longer vanishingly small, the dynamics are significantly modified, with a more gradual decay in the coherence. This behaviour is expected to become relevant for relatively small ensembles outside the scale-separated regime.

At later times, $t\gtrsim \sigma/g^2$, the highly oscillatory region grows to cover most of the Gaussian TLF distribution, so that only the tails (rare configurations) contribute appreciably. As these contributions evolve, their phases can partially re-align, producing constructive interference and leading to partial revivals of the oscillator’s coherence.
At asymptotically large times, $t\gg\sigma/g^2$, the coherence is governed only by the extreme tails of the distribution. However, at these times the signal has already been strongly suppressed by earlier phase averaging, rendering this timescale largely irrelevant.

\section{Conclusions and discussion}
\label{section: conclusions and discussion}

We have used a simple theoretical model to describe the effects of a nearly resonant TLS coupled to one or more TLFs on the coherence of an oscillator. The oscillator and TLS form a JC system, displaying Rabi oscillations which modulate the oscillator coherence. The TLFs add additional coherence oscillations which can either be superimposed on or suppress the Rabi oscillations. Bath-induced transitions in the TLF states lead to an irreversible loss of coherence, but the frequency mixing arising from even a relatively small set of TLFs is often sufficient to rapidly degrade the coherence.

Motivated by recent experiments on phononic crystal resonators---which suggest that the number of TLFs involved may be small\,\cite{maksymowych2025,yuksel2025}---we initially considered the coherence dynamics induced by a single TLF coupled to the TLS. When the TLF is weakly coupled, it leads to a further periodic oscillation superimposed on top of the Rabi oscillations, with dissipation leading to an exponential decay. In contrast, a strongly-coupled TLF can wash out the Rabi oscillations almost completely, instead leading to strong and slow periodic oscillations in the oscillator coherence with frequency $g^2/2|\lambda|$. Dissipation again causes the oscillations to decay exponentially, though the rate depends sensitively on the specific parameter regime. Surprisingly, if the dissipation is strong enough, Rabi oscillations can re-emerge temporarily as the coherent effect of the TLFs is attenuated.

We also considered the effects of an ensemble of $N$ non-dissipative TLFs, deriving analytic approximations that describe the behaviour in the limits of a narrow or broad (compared to the JC coupling $g$) TLF distribution for large dense ensembles by relying on the CLT. Numerical calculations show that the analytic expressions can be surprisingly accurate for relatively small sets of TLFs, especially during the initial stages of the evolution. Situations where several TLFs are present, but one couples much more strongly than the rest, violate the requirements of the CLT and deviate in important ways from the large-ensemble limit. Such systems tend to display stronger oscillations in the coherence, less frequency mixing and a slower overall decay.

Our work focused on the simplest possible superposition state for the oscillator (a superposition of the $|0\rangle$ and $|1\rangle$ Fock states) throughout. The obvious next step is to go beyond this to fully explore the impact of TLS--TLF-induced decoherence on the wide range of oscillator superposition states that can be produced (including coherent
states and cat states).
However, our focus on the reduced problem of coherence in a two-state basis also allows the approach to be applied to qubit--TLS systems, where near-resonant couplings and TLF-induced dephasing are both present\,\cite{Matityahu2024}. 

Finally, we note that extending the relatively simple model system studied here to  include more elements will be important to make it more broadly relevant to situations encountered in experiments. Some of the most important extensions include adapting the local master equation approach we used to account for dissipation with a single TLF to an ensemble of TLFs, allowing for direct oscillator--TLF couplings and including additional loss channels (for example, dissipation of the TLS due to phonons). Including dissipation for the TLF ensemble would reduce the need to rely on assumptions about how fast coherence decays, and including other dissipation channels would likely give rise to additional exponential decay contributions, potentially reducing the visibility of features like coherence oscillations.

\section*{Data Availability}

The data supporting the findings of this study are openly available in the Nottingham Research Data Management Repository at \href{https://doi.org/10.17639/nott.18008}{DOI: 10.17639/nott.18008}.

\ack
    
We thank Miles P. Blencowe  for very helpful discussions. TJA acknowledges support from EPSRC (UKRI, UK) in the form of a studentship (EP/W524402/1). JG and ADA acknowledge support from the Leverhulme Trust under Research Project Grant Ultra-Cool Mechanics (RPG-2023-177).

\appendix

\section{Connection to microscopic properties of TLSs}
\label{appendix: connection to microscopic properties of the ensemble}

Our approach is based on the study of a simplified theoretical model, rather than a specific physical realisation of the oscillator--TLS--TLF system. Nevertheless, the microscopic details of TLSs in quantum devices have inspired our model building. Therefore, in this appendix we give a brief summary of the microscopic details that provide the context for the main text, and the different regimes we chose to focus on. We start by reviewing the properties of TLSs, before going on to derive an expression for the coupling variance in the continuum limit using standard assumptions.  

\subsection{TLS properties}
\label{app: TLS properties}

The properties of TLSs in engineered quantum systems can depend strongly on the materials involved, the size and shape of the system and the temperature of the surroundings\,\cite{behunin2016,muller2019}.
An individual TLS is described by a Hamiltonian of the form
\begin{equation}
    \hat{H}_{\rm{TLS}} = \frac{1}{2}\left(\Delta\hat{\tilde{\sigma}}_z+\Delta_0\hat{\tilde{\sigma}}_x \right),
\end{equation}
with $\Delta$ the asymmetry, $\Delta_0$ the tunnel splitting and $\hat{\tilde{\sigma}}_{x,z}$ the Pauli operators in a physical basis\,\footnote{Physically, TLSs are generally thought to consist of atomic defects tunnelling between different locations within a disordered lattice, and the basis states correspond to the minima of a double-well potential\,\cite{phillips1987}.}. In amorphous solids, the distribution of TLS parameters is typically\,\cite{phillips1987,stephens2021,matityahu2016,carruzzo2020} assumed to be described by a probability density of the form $P(\Delta,\Delta_0)=P_0/\Delta_0$, with $P_0$ a constant. However, there remains considerable uncertainty about whether such a distribution describes the oxide layers and disordered surfaces where TLSs are thought to reside in engineered quantum devices\,\cite{burnett2014,faoro2015,behunin2016}.

Within insulating systems, TLS--phonon coupling controls the relaxation and dissipation into the thermal bath. Standard perturbation methods lead to a phonon-mediated relaxation rate\,\cite{behunin2016,matityahu2016} 
\begin{equation}
    \gamma_1(\varepsilon,\Delta_0) = \chi\varepsilon^{d-2}\Delta_0^2\,\coth\left(\frac{\varepsilon}{2k_{\rm{B}}T}\right),
\end{equation}    
where $\varepsilon=\sqrt{\Delta^2+\Delta_0^2}$, $\chi$ is a constant collecting together material-dependent parameters, $d$ is the dimensionality of the host medium and $T$ the temperature. The relaxation rate is maximised for cases where $\Delta_0=\varepsilon$, and at the border of the thermally activated regime with $\varepsilon\sim k_{B}T$ we have $\gamma_1^{\rm{max}}\sim T^d$. At sufficiently low temperatures, the relaxation of even the fastest relaxing TLSs can therefore become very slow. Thus, several experiments have been able to access the regime where pure TLF dephasing dominates over phonon-mediated relaxation\,\cite{lisenfeld2016,matityahu2016,hitchcock26}. In phononic crystal structures, a phonon band-gap engineered to greatly suppress coupling of the mechanical mode to the surrounding phonons will also strongly reduce the relaxation rate of near-resonant TLSs\,\cite{bozkurt2025,maksymowych2025,yuksel2025}.   

Electric and elastic dipolar couplings mediate an effective TLS--TLS interaction of the form\,\cite{remus2012,carruzzo2020,matityahu2016,maksymowych2025} $\hat{H}_{{\rm{int}}}= \sum_j\sum_{i\neq j}(J_{ij}/2)\hat{\tilde{\sigma}}_{z,i}\hat{\tilde{\sigma}}_{z,j}$. When rotating into the eigenstate basis and considering TLS--TLF couplings in particular---where the energy scales are very different---one finds that only the dispersive interaction needs to be retained\,\cite{muller2015, faoro2015, matityahu2016, maksymowych2025}. This leads to an interaction of the form used in the main text, equation (\ref{equation: hamiltonian}),
\begin{equation}
    \hat{H}_{\rm{TLS-TLF}}=\sum_{j}\lambda_j\hat{{\sigma}}_{z}\hat{{\tau}}_{z,j},
\end{equation}    
where\,\footnote{Note that we have set aside signs arising from the orientation of the dipoles; the distribution of couplings is expected to be even in the continuum limit leading to vanishing mean\,\cite{carruzzo2020,muller2015}.}
\begin{equation}
    \lambda_j^2=J_0^2\left(\frac{r_0}{r_j}\right)^{2d}\Biggl(\frac{\Delta_T}{\varepsilon_T}\Biggr)^2\left(\frac{\Delta_j}{\varepsilon_j}\right)^2,
\end{equation}
with $J_0$ a material-dependent constant, $r_0$ a short-length cut-off, $\Delta_{T(j)}$ the TLS (TLF) asymmetry and $r_j$ the TLS--TLF distance. The strong dependence on position means that unless the density is sufficiently high,  meeting the conditions of the CLT may be difficult and, indeed, the opposite limit of a single TLF can become the relevant one\,\cite{lisenfeld2016,matityahu2016,maksymowych2025,yuksel2025}. The coupling strengths, $\lambda_j$, will necessarily vary considerably, but values of order $100$\,MHz have been reported\,\cite{lisenfeld2015}. In contrast, recent experiments\,\cite{maksymowych2025,yuksel2025} on phononic crystal devices have estimated rather smaller oscillator--TLS couplings, $g/2\pi\sim 1-2$\,MHz. Furthermore, frequency fluctuations reported in Ref.\,\cite{maksymowych2025} were found to be consistent with an oscillator--TLS coupling and a TLS--TLF coupling (to a single TLF) deep inside the strong-coupling regime, $\lambda/g\sim 200$.

\subsection{Continuum limit calculation of coupling variance}
\label{app: TLS variance}
Within the continuum limit, we can perform an average over the TLF ensemble to calculate the average variance $\braket{\sigma^2}$ and hence estimate the average oscillator coherence decay rate (see section \ref{section: ensemble of fluctuators}). This provides a way to obtain the dependencies on temperature and the parameters of the TLS. Following the approach in Ref.\,\cite{matityahu2016}, we obtain
\begin{equation}
    \braket{\sigma^2} = \int d\Omega_{d} \int^\infty_{r_0} r^{d-1}dr \int^\infty_0 d\varepsilon \int^1_{u_\text{min}}du\,P(\varepsilon,u)\,\lambda^2(r,u)\sech^2\left(\frac{\varepsilon}{2k_{\rm{B}}T}\right),
\end{equation}
where $\Omega_d$ is the $d$-dimensional solid angle and we have expressed the TLF distribution in terms of $\varepsilon$ and $u=(\Delta_0/\varepsilon)^2$, $P(\varepsilon,u)=P_0/(2u\sqrt{1-u})$, with a cut-off $u_{\rm{min}}$ assumed to avoid divergence\,\cite{stephens2021}. This uniform-in-$\varepsilon$ distribution leads  to a linear temperature dependence and we find that, for both $d=2$ and $d=3$,
\begin{equation}
    {\braket{\sigma^2}} \propto J_0^2P_0\,k_{\rm{B}}T\cos^2\theta,
\end{equation}
with $\cos\theta=\Delta_T/\varepsilon_T$, determined by the specific TLS parameters.

\section{Higher-order approximate solution for a strongly coupled fluctuator}
\label{appendix: higher-order approximate solution for a single non-dissipating, strongly-coupled fluctuator}

In this appendix, we briefly give the result obtained when including higher-order terms, $\mathcal{O}(g^2/\lambda^2)$, in our approximate solution for the case of a single non-dissipating, strongly-coupled TLF. In the limit $|\lambda|\gg g$ and assuming $\delta = 0$, we can rewrite our exact analytical solution, equation (\ref{equation: single tlf jc coherence}), as
\begin{equation}
    C(t) \simeq \sqrt{\left(A\cos\left(\frac{g^2t}{2\lambda}\right) + B\cos2\lambda t\right)^2 + \tanh^2\left(\frac{\varepsilon}{2k_{\rm{B}}T}\right)\left(A\sin\left(\frac{g^2t}{2\lambda}\right) + B\sin2\lambda t\right)^2},
\label{equation: strongly coupled single tlf approximation at higher order}
\end{equation}
where $A = 1 - B$ and $B = g^2/4\lambda^2$. Here, the terms oscillating with frequency $g^2/2|\lambda|$ set the dominant slow envelope behaviour, while the terms oscillating with frequency $2|\lambda|$ set the small, fast Rabi oscillations. If we ignore the effect of the Rabi oscillations, and focus on the leading behaviour by taking $A \approx 1$ and $B \approx 0$, we recover the result presented in the main text, equation (\ref{equation: strongly coupled single tlf approximation at leading order}).

\section{Onset of continuum-like behaviour in narrow ensembles}
\label{appendix: statistical analysis of the onset of continuum-like behaviour in narrow ensembles}

In this appendix, we provide a more detailed statistical analysis of how and when coherence dynamics associated with a `large' and narrow ($g\gg\sigma$) ensemble of TLFs emerge, characterised by continuum-like behaviour well described by the CLT. In particular, we investigate how agreement between the approximate continuum solution and an exact calculation depends on the number of TLFs, $N$, and the dominance of individual TLF contributions, as quantified by $R=\max_j(\sigma_j^2)/\sigma^2$.

\begin{figure}[t!]
\centering
\includegraphics[width = 0.96\linewidth]{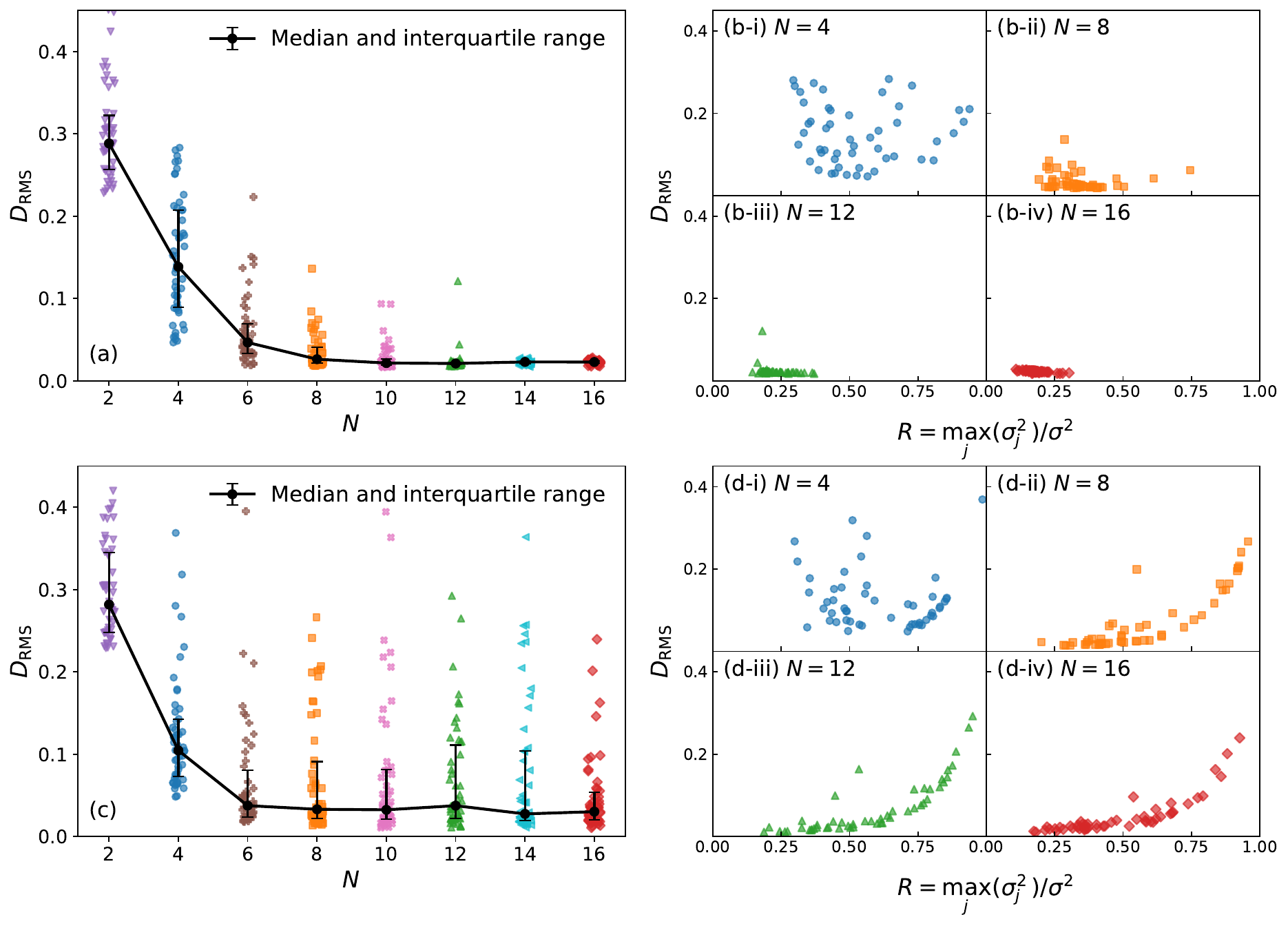}
\caption{Behaviour of $D_{\rm{RMS}}$ as a function of the TLF ensemble size, $N$, and the largest individual TLF contribution to the total variance, $R=\max_j(\sigma_j^2)/\sigma^2$. Each ensemble size is sampled over 50 independent coupling realisations, with deviations evaluated over $0\leq\sigma t/\pi\leq3$. Panels (a,b) are for a uniform-random distribution (parameters matching those in figure \ref{figure 4}), whilst panels (c,d) are for location-random distributions (parameters matching those in figure \ref{figure 5}).  Values of $D_{\rm{RMS}}$ for individual realisations (coloured markers), together with the median and interquartile range (black), are shown in (a,c) as a function of $N$. Panels (b-i)--(b-iv) and (d-i)--(d-iv) show the dependence of $D_\text{RMS}$ on $R$ for $N=4,8,12$, and $16$, respectively.}
\label{figure C1}
\end{figure}

We quantify the difference between the exact and continuum solutions using the root-mean-square (RMS) deviation,
\begin{equation}
    D_\text{RMS} = \sqrt{\frac{1}{M}\sum_{k=1}^M\left[C(t_k) - C_\text{cont}(t_k)\right]^2},
\end{equation}
where $C(t)$ is the (numerically evaluated) exact coherence [equation (\ref{equation: many tlf jc coherence})], and $C_\text{cont}(t)$ is the continuum solution obtained under the assumptions of a narrow ensemble and the applicability of the CLT [equation (\ref{equation: narrow tlf short time approximation})]. Note that the RMS deviation is evaluated numerically at $M=500$ uniformly spaced time points in the interval $0\leq \sigma t/\pi\leq 3$. Results for the uniform-random and location-random coupling distributions sampled in the main text (see figures~\ref{figure 4} and \ref{figure 5}, respectively) are presented in figure~\ref{figure C1}, with parameter values chosen to match those used in the corresponding main-text figures.

Figures~\ref{figure C1}(a,c) show that increasing $N$ generally decreases the median RMS deviation, indicating improved agreement between the exact finite-$N$ solution and the continuum approximation. For both coupling distributions, the largest reduction occurs at relatively small $N$, with continuum-like behaviour beginning to emerge for ensembles with only a few TLFs. Beyond this initial reduction, the median deviation changes relatively little with further increases in ensemble size ($N>8$), especially for uniform-random couplings, suggesting that the continuum approximation already provides a good description of the typical coherence dynamics for modest ensemble sizes ($N$ in the range $6\text{--}8$). However, the spread in deviations differs considerably between the two types of distribution. For uniform-random couplings, the deviations become increasingly concentrated around a small value as $N$ increases, whereas location-random couplings continue to exhibit significant sample-to-sample variation up to $N=16$. This demonstrates that increasing the number of TLFs generally favours continuum-like behaviour, but does not guarantee its emergence for a particular coupling realisation.

Figures~\ref{figure C1}(b-i)--(b-iv) and~\ref{figure C1}(d-i)--(d-iv) show how the RMS deviation depends on the largest individual TLF contribution to the total variance, quantified by $R$. In general, larger values of $R$ are associated with greater deviations from the continuum approximation, indicating that the dominance of one TLF (or potentially a few TLFs) can inhibit the emergence of continuum-like behaviour. This is most important for location-random couplings, where substantial differences in individual coupling strengths can lead to large values of $R$ even for relatively large ensembles. Conversely, smaller values of $R$ generally favour agreement with the continuum approximation, although a small $R$ alone does not guarantee continuum-like behaviour, particularly for small $N$, where the finite number of TLF configurations remains important. Indeed, some realisations exhibit comparatively large deviations despite relatively small values of $R$. Such cases may arise when a small subset of TLFs collectively dominates the total variance, even though no single TLF makes an overwhelmingly large contribution. This highlights a limitation of $R$ in characterising the collective contributions of multiple TLFs. These results demonstrate that the onset of continuum-like behaviour depends not only on the ensemble size, but also on how evenly the total variance is distributed among the constituent TLFs.

\section{Dissipative dynamics from a local master equation}
\label{appendix:dissipation}
We take the effect of coupling between a TLF and the thermal bath into account phenomenologically using a `local' Lindblad master equation\,\cite{cattaneo2019}
\begin{equation}
\frac{d\hat{\varrho}}{dt}=-i[\hat{H},\hat{\varrho}]+\frac{\gamma}{2}\left(2\hat{\tau}_-\hat{\varrho}\hat{\tau}_+-\{\hat{\tau}_+\hat{\tau}_-,\hat{\varrho}\}\right)+\frac{\gamma}{2}\left(2\hat{\tau}_+\hat{\varrho}\hat{\tau}_- -\{\hat{\tau}_-\hat{\tau}_+,\hat{\varrho}\}\right), 
\label{equation: master equation}
\end{equation}
where $\hat{\tau}_{\pm}$ are the TLF raising and lowering operators. We have assumed the scale-separated limit so that the upward and downward rates for the TLF are the same, given by $\gamma$. For simplicity, we  will also assume from now on that the TLS and oscillator are resonant, $\delta=0$. 

This Lindblad equation is readily justified in the regime where the TLF is only very weakly coupled to the TLS and oscillator, $|\lambda|\ll g$. However, it can also be applied in the opposite regime where $g\ll |\lambda|$. In this regime, the `local' approximation neglects the oscillator and treats the TLS and TLF together as the system coupled to the bath, but the simple dispersive coupling ensures that it always remains diagonal in the $\hat{\tau}_z$, $\hat{\sigma}_z$ basis. Provided $|\lambda|\ll k_{\rm{B}}T\ll \varepsilon_T$, the TLS--TLF eigenstates consist of a pair of doublets with intra-doublet level spacings $|{\varepsilon}\pm2\lambda|\ll k_{\rm{B}}T$, and the bath produces intra-doublet transitions with rates that can be approximated as equal, provided the density of states of the bath is sufficiently slowly varying.

For the simple initial state we consider here, equation (\ref{equation: jc initial state}), the coherence can be obtained from $\langle \hat{\mathcal{O}}_{\pm}(t)\rangle$ where $\hat{\mathcal{O}}_\pm = \ket{\mathcal{G}}\bra{1\pm}$ are the two lowest lowering operators within the JC eigenbasis. The evolution of these expectation values, together with $\langle \hat{\tau}_z\hat{\mathcal{O}}_{\pm}(t)\rangle$, forms a set of four coupled equations
\begin{eqnarray}
\frac{d\langle \hat{\mathcal{O}}_{\pm}(t)\rangle}{dt}&=&-i\Delta\omega_{\pm}\langle \hat{\mathcal{O}}_{\pm}(t)\rangle-i\lambda\left[\langle\hat{\tau}_z \hat{\mathcal{O}}_{\pm}(t)\rangle-\langle \hat{\tau}_z\hat{\mathcal{O}}_{\mp}(t)\rangle\right],\\
\frac{d\langle {\hat{\tau}_z\hat{\mathcal{O}}}_{\pm}(t)\rangle}{dt}&=&-\left[i\Delta\omega_{\pm}+2\gamma\right]\langle \hat{\tau}_z\hat{\mathcal{O}}_{\pm}(t)\rangle-i\lambda\left[\langle\hat{\mathcal{O}}_{\pm}(t)\rangle-\langle\hat{\mathcal{O}}_{\mp}(t)\rangle\right],
\end{eqnarray}
where $\Delta\omega_{\pm}=\omega_0\pm g$ are the frequencies of the transitions between the first JC doublet states and the ground state.

\subsection{Weakly coupled fluctuator: $|\lambda| \ll g$}

For a weakly coupled TLF, we can proceed by moving to a rotating frame to eliminate the common frequency $\omega_0$, and then making a rotating wave approximation, neglecting the counter-rotating terms which couple the two pairs of equations,
\begin{eqnarray}
\frac{d\langle \hat{{\mathcal{O}}'}_{\pm}(t)\rangle}{dt}&=&\mp ig\langle \hat{\mathcal{O}}'_{\pm}(t)\rangle-i\lambda\langle\hat{\tau}_z \hat{\mathcal{O}'}_{\pm}(t)\rangle,\\
\frac{d\langle {\hat{\tau}_z\hat{\mathcal{O}}}'_{\pm}(t)\rangle}{dt}&=&-\left[\pm ig+2\gamma\right]\langle \hat{\tau}_z\hat{\mathcal{O}'}_{\pm}(t)\rangle-i\lambda\langle\hat{\mathcal{O'}}_{\pm}(t)\rangle,
\end{eqnarray}
where the prime indicates the frame rotating at frequency $\omega_0$ (i.e. $\hat{\mathcal{O}}'_{\pm}=\hat{\mathcal{O}}_{\pm}{\rm{e}}^{i\omega_0t}$). Solving these coupled equations then leads directly to a second-order equation in $\langle \hat{{\mathcal{O}}'}_{\pm}(t)\rangle$, whose solution is equivalent to damped harmonic motion, giving rise to the expression for $C(t)$ in the main text, equation (\ref{eq:wcfdecay}).

\subsection{Strongly coupled fluctuator: $|\lambda|\gg g$}

A different set of approximations is needed for a strongly coupled TLF, and it is convenient to define a set of sum and difference variables:
\begin{eqnarray}
X_\pm&=&\langle \hat{{\mathcal{O}}'}_{+}(t)\rangle\pm\langle \hat{{\mathcal{O}}'}_{-}(t)\rangle,\\
Y_\pm&=&\langle \hat{\tau}_z\hat{{\mathcal{O}}'}_{+}(t)\rangle\pm\langle \hat{\tau}_z\hat{{\mathcal{O}}'}_{-}(t)\rangle,
\end{eqnarray}
which evolve according to
\begin{eqnarray}
\dot{X}_+&=&-igX_- \label{eq:x+},\\
\dot{X}_-&=&-igX_+-i2\lambda Y_-\label{eq:x-},\\
\dot{Y}_+&=&-igY_--2\gamma Y_+\label{eq:y+},\\
\dot{Y}_-&=&-igY_+-2\gamma Y_--i2\lambda X_- \label{eq:y-}.
\end{eqnarray}
Note that the coherence is given by $C(t)=\sqrt{2}|X_+(t)|$, and that we have the initial conditions $X_+(0)=1/\sqrt{2}$ and $X_-(0)=Y_{\pm}(0)=0$. Hence, the coherence decay becomes slower as $g$ decreases and vanishes in the limit $g\rightarrow 0$.
Depending on the relative size of $\gamma $, three distinct regimes can be identified.

\subsubsection{Strongly damped, $\gamma\gg |\lambda|$:}
When the damping is strong enough, we can make use of adiabatic elimination. Hence, we have $Y_+=-igY_-/2\gamma$ and $Y_-=-i(\lambda X_-/\gamma)/(1-g^2/4\gamma^2)\simeq -i\lambda X_-/\gamma$. This leads to damped harmonic motion for $X_+(t)$, with solution 
\begin{equation}
X_+(t)=\frac{{\rm e}^{-\Gamma t}}{\sqrt{2}}\left(\cos\kappa t+\frac{\Gamma}{\kappa}\sin\kappa t\right),    \label{eq:dho}
\end{equation}
where $\kappa=\sqrt{g^2-(\lambda^2/\gamma)^2}$ and $\Gamma=\lambda^2/\gamma$. 
\subsubsection{Weakly damped, $\gamma\ll |\lambda|$:}
In the weakly damped regime, the separation of frequencies, $|\lambda|\gg g$, means that $X_+,Y_+$ evolve much more slowly than $X_-,Y_-$. We can simplify by adopting new sum and difference variables using the slowly and rapidly evolving quantities,
$a_{\pm}= X_-\pm Y_-$ and $b_{\pm}= X_+\pm Y_+$.
Neglecting damping for now, and moving to the fast rotating frame, $\tilde{a}_{\pm}=a_{\pm}{\rm{e}}^{\pm i2\lambda t}$, we find
\begin{equation}
\dot{\tilde{a}}_{\pm}=-igb_{\pm}{\rm{e}}^{\pm i2\lambda t}.
\end{equation}
To lowest order in $g$, we can treat $b_{\pm}$ as constant, so that $a_{\pm}\simeq \mp\frac{g}{2\lambda}b_{\pm}$. Including damping at this stage produces a sub-leading correction, as we have assumed $|\lambda|\gg \gamma, g$.

Returning to the $X_{\pm}, Y_{\pm}$ variables, our approximation implies that $X_-(Y_-)\simeq -\frac{g}{2\lambda}Y_+(X_+)$. Substituting these into equations (\ref{eq:x+}) and (\ref{eq:y+}), differentiating the equation for $\dot{X}_+$, and eliminating $Y_+$, we recover equation (\ref{eq:dho}) with $\kappa=\sqrt{(g^2/2\lambda)^2-\gamma^2}$ and $\Gamma=\gamma$.

\subsubsection{Intermediate damping, $\gamma \sim |\lambda|$:}
In the intermediate regime, we can  neglect the effect of $Y_+$ entirely\,\footnote{Treating $Y_+$ adiabatically as $\gamma\gg g$, we find $Y_+\simeq -igY_-/2\gamma$. But substituting this into equation (\ref{eq:y-}) only leads to a correction of order $(g/\gamma)^2$ which can be dropped.}. The equations of motion for $X_{\pm}$ and $Y_-$ can then be combined into the third-order equation,
\begin{equation}
\dddot{X}_++2\gamma\ddot{X}_++(g^2+4\lambda^2)\dot{X}_++2\gamma g^2X_{+}=0.
\end{equation}

We seek solutions of the form $X\sim {\rm{e}}^{x t}$, for which  a cubic equation follows immediately 
\begin{equation}
x^3+2\gamma x^2+(g^2+4\lambda^2)x+2\gamma g^2=0.
\end{equation}
In the limit $g\rightarrow 0$, the three roots consist of a zero mode and two fast-decaying ones. The coherence dynamics is controlled by the $g$-dependent slow mode that evolves from the $x=0$ solution when  $g> 0$. Since the relevant root is small, we can obtain it approximately by neglecting the $x^3$ term and solving the corresponding quadratic equation
\begin{equation}
    x=-\lambda^2/\gamma\pm\sqrt{(\lambda^2/\gamma)^2-g^2}.
\end{equation}
Using $g\ll \gamma, |\lambda|$, we find the slow  ($g$-limited) root in this case and hence obtain $X_{+}(t)\simeq{\rm{e}}^{-g^2\gamma t/2\lambda^2}/\sqrt{2}$.

\newcommand{\newblock}{}
\bibliographystyle{iopart-num.bst}
\bibliography{references/references}

\end{document}